\documentclass[twocolumn,referee]{aastex6}
\usepackage{graphicx,amsmath}
\usepackage[varg]{txfonts}
\usepackage{booktabs,makecell}
\usepackage{natbib}
\usepackage{multirow}

\begin{document}

\title{A method to measure the transverse magnetic field and\\ orient the rotational axis of stars}
\author{Francesco Leone, Cesare Scalia, Manuele Gangi, Marina Giarrusso,}
\affil{Universit\`a di Catania, Dipartimento di Fisica e Astronomia, Sezione Astrofisica,  \\
       Via S. Sofia 78, I--95123 Catania, Italy}
\author{Matteo Munari, Salvatore Scuderi, Corrado Trigilio}
\affil{INAF - Osservatorio Astrofisico di Catania, Via S. Sofia 78, I--95123 Catania, Italy}
\author{Martin Stift}
\affil{Armagh Observatory, College Hill, Armagh BT61 9DG. Northern Ireland}

\begin{abstract}
Direct measurements of the stellar magnetic fields are based on the splitting of spectral
lines into polarized Zeeman components. With few exceptions, Zeeman signatures are hidden
in data noise and a number of methods have been developed to measure the average, over the
visible stellar disk, of longitudinal components of the magnetic field. As to faint stars,
at present observable only with low resolution spectropolarimetry, a method is based on the
regression of the Stokes\,$V$ signal against the first derivative of Stokes\,$I$.
Here we present an extension of this method to obtain a direct measurement of the
transverse component of stellar magnetic fields by the regression of high resolution
Stokes\,$Q$ and $U$ as a function of the second derivative of Stokes\,$I$.
We also show that it is possible to determine the orientation in the sky of the rotation
axis of a star on the basis of the periodic variability of the transverse component due
to its rotation. The method is applied to data, obtained with the {\it Catania
Astrophysical Observatory Spectropolarimeter}, along the rotational period of the
well known magnetic star $\beta$\,CrB.
\end{abstract}


\section{Introduction}

In stellar astrophysics, magnetic fields are measured by means of the Zeeman effect,
whereby the $(2\,J + 1)$-fold degeneracy of the fine structure levels of the various
spectroscopic terms is completely lifted by a magnetic field. This results in the
splitting of a spectral line into Zeeman components: the $\sigma_{-}$- and
$\sigma_{+}$-components ($\Delta M = \pm 1$) are circularly polarized, the 
$\pi$-components ($\Delta M = 0$) linearly.
For weak to moderate fields, the displacements in wavelength of the $\sigma$-components
from the unsplit line position $\lambda_{0}$ (in {\AA}) due to a magnetic field $\vec{B}$
(in G) is given by
\begin{equation}
   \Delta \lambda = 4.67\,10^{-13}\,\bar{g}\,\lambda_{0}^2\,|\vec{B}|\label{Dl}
\end{equation}
where $\bar{g}$ is the so called ``effective Land{\'e} factor'', related to the
Land{\'e} factors $g_1$ and $g_2$ of the involved energy levels by
\begin{align} \label{foG}
   \bar{g}       & = 0.5\,(g_{1} + g_{2}) + 0.25\,(g_{1} - g_{2})\,d\\
   {\rm with}~~d & = [J_{1} (J_{1} + 1) - J_{2} (J_{2} + 1)]\nonumber
\end{align}

{From Eq.\,\ref{Dl} it transpires that resolved Zeeman components can rarely be
observed in optical spectra.} To give an example, the $\sigma_{-}$ and the $\sigma_{+}$
components of a simple Zeeman triplet ($\bar{g} = 1.0$) at $\lambda$ = 5000\,{\AA},
split in a 1\,kG magnetic field, would overlap for a projected rotational velocity
$v_{\rm e}\,\sin i \approx 1.5$\,km\,s$^{-1}$ or an instrumental resolution of
$R = 200\,000$. In order to establish the presence of a stellar magnetic field,
it rather makes sense to measure the distance between the respective centers of
gravity of a spectral line in left-hand (lcp) and right-hand (rcp) circularly
polarized light. The distance in wavelength between the lcp and rcp centers of
gravity is proportional to the disk-averaged line-of-sight component $B_{\rm z}$
of the magnetic field vector, called ``effective magnetic field'' by \cite{Babcock47}.
\begin{equation}
B_{||} = \frac{1}{W \cal F_{\it I_{\rm c}}} \int_{0}^{2\pi}d\phi
   \int_{0}^{\pi/2}B_{\rm z}\cos\theta\,\sin\theta\,d\theta \times
   \int [I_{\rm c} - I_\lambda]\,d\lambda\label{Beff}
\end{equation}
where $W$ is the equivalent width of the line, ${\cal F_{\it I_{\rm c}}}$ denotes the
continuum flux at the wavelength of the line; $\phi$ and $\theta$ are polar
coordinates. $I_{\rm c}$ and $I_\lambda$ represent the respective continuum and line
intensities at the coordinate ($\theta$, $\phi$). 
$B_{||}$ is commonly obtained from the relation given by \cite{Mathys1994}:
\begin{equation}
R_V^{(1)} = \frac{1}{W}\int{\frac{V_c - V_{\lambda}}{I_c}\,(\lambda-\lambda_0)\,d\lambda}
        = 4.67\,10^{-13}\,\bar{g}\,\lambda_0^2\,B_{||} \label{RV}
\end{equation}

It is a fact that with increasing instrumental smearing, Stokes polarization profiles
rapidly become unobservable  \citep{Leone01, Leone03}; on the other hand, high
resolution spectropolarimetry is at present limited to bright (V$\lesssim 10$) stars. To
overcome these limitations, \cite{Angel70} introduced a method based on narrow-band
($\sim$ 30\,\AA) circular photopolarimetry in the wings of Balmer lines for the
measurement of magnetic fields of stars that could not be observed with high-resolution
spectropolarimetry. The difference between the opposite circularly polarized
photometric intensities is converted to a wavelength shift and subsequently to
the effective longitudinal field ${\rm{B_{||}}}$. Another method, suggested by
\cite{Bagnulo02}, is based on the relation between Stokes\,$V$ and $I$ for spectral
lines whose intrinsic width is larger than the magnetic splitting \citep{Mathys89}:
\begin{equation}
  \frac{V_\lambda}{I_\lambda} =-4.67\,10^{-13}\,\bar{g}\,
\lambda^{2}\,B_{||}\,\frac{1}{I_{\lambda}}\,\frac{\partial I_\lambda}{\partial \lambda} \label{gV}
\end{equation}
This linear fitting of Stokes\,$V$ against the gradient of Stokes\,$I$
(Eq.\,\ref{gV}) to measure the effective magnetic field of faint targets on the
basis of low resolution spectropolarimetry without wasting any circular polarized
signal has opened a new window. A method to measure the magnetic fields of
previously inscrutable objects has indeed been largely used. The reader can refer
to \cite{Bagnulo15} for a review on this method and its results.

The problem of measuring the magnetic field of faint stars represents a special case
of the more general problem of how to recover Stokes profiles ``hidden'' in photon noise. 
With reference to the very weak magnetic fields of late-type stars, the solution introduced 
by a lamented colleague and friend, Meir Semel, consisted in adding the Stokes\,$V$
profiles of all lines present in a spectrum, obtaining a {\it pseudo profile} of
a very high signal to noise (S/N) ratio \citep{Semel96}. This idea has been further
developed by \cite{Donati97} who introduced the {\it Least Squares Deconvolution}
(LSD) method. Later, \cite{Semel06} initiated yet another approach to the add-up
of Stokes profiles from noisy spectra, based on {\it Principal Component Analysis}.

The measurement of the $B_{||}$ component is important to assign a lower limit to the
strength of a magnetic field. But in order to constrain the magnetic topology, the
transverse component $B_{\bot}$ is necessary too. To our knowledge, no direct measurements
of the transverse component of a stellar magnetic field have yet been obtained. 
No relations similar to Eqs.\,\ref{RV} and \ref{gV} have yet been implemented.
According to \cite{Landi04}, Stokes\,$Q$ and $U$ are related to the second derivative
of Stokes\,$I$ by
\begin{align}
  \frac{Q_\lambda}{I_\lambda} = & -5.45\,10^{-26}\,\bar{G}\,
\lambda^{4}\,B_{\bot}^2\,\cos\,2\chi\,\frac{1}{I_{\lambda}}\,\frac{\partial^2 I}{\partial\lambda^2}\label{gQ}\\
  \frac{U_\lambda}{I_\lambda} = & -5.45\,10^{-26}\,\bar{G}\,
\lambda^{4}\,B_{\bot}^2\,\sin\,2\chi\,\frac{1}{I_{\lambda}}\,\frac{\partial^2 I}{\partial\lambda^2}\label{gU}
\end{align}
where
\begin{equation}
    \bar{G} = \bar{g}^2 - \delta\label{soG}
\end{equation}
is the second order effective Land{\'e} factor, with
\begin{align*}
   \delta & = (g_1 - g_2)^2\,(16 s - 7d^2 -4)/80\\
   s      & = [J_1 (J_1 + 1) + J_2 (J_2 + 1)]
\end{align*}
and $J_1$ and $J_2$ the angular momenta of the involved energy levels.

Stokes\,$Q$ and $U$ signals across the line profiles are weaker than the $V$ signal
and instrumental smearing is more destructive for Stokes\,$Q$ and $U$ profiles than
for Stokes $V$ \citep{Leone03} because their variations are more complex and
occur on shorter wavelength scales. As a result, Stokes\,$Q$ and $U$ have rarely
been detected -- being hidden in the noise even in stars characterized by very
strong Stokes $V$ signals -- but it is worth mentioning that \cite{Wade2000} have
successfully applied the LSD method also to Stokes\,$Q$ and $U$ profiles. When
observed, Stokes\,$Q$ and $U$ profiles represent a strong constraint to the magnetic
geometry \citep{Bagnulo01}. Following \cite{Landi1981} who showed that broadband
linear polarization arises from saturation effects in spectral lines formed in a
magnetic field \citep{Calamai1975}, \cite{Bagnulo1995} have used phase-resolved
broadband linear photopolarimetry to constrain stellar magnetic geometries.

In Section\,\ref{Sec_Reg}, we show that application of the linear regression method
to high resolution Stokes\,$V$ spectra results in highly accurate measurements of
the stellar effective magnetic field (hereafter {\it longitudinal field}). An
extension of this regression method to high resolution Stokes\,$Q$ and $U$ spectra
on the other hand results in a direct measure of the mean transverse component of
the field (hereafter {\it transverse field}).
For this purpose, we have obtained a series of full Stokes $IQUV$ spectra of
$\beta$\,CrB (Section\,\ref{Sec_Obs}) over its rotational period with the
{\it Catania Astrophysical Observatory Spectropolarimeter} \citep{Leone16}.

In Section\,\ref{Sec_Orient}, we will show that, as a consequence of the
stellar rotation, the transverse component of the magnetic field describes
a closed loop in the sky, offering the possibility to determine the
orientation of the rotational axis.

\begin{figure*}
\centering
\includegraphics[trim = 0.5cm 1.0cm 0.0cm 0.0cm, clip=true,width=5.9cm]{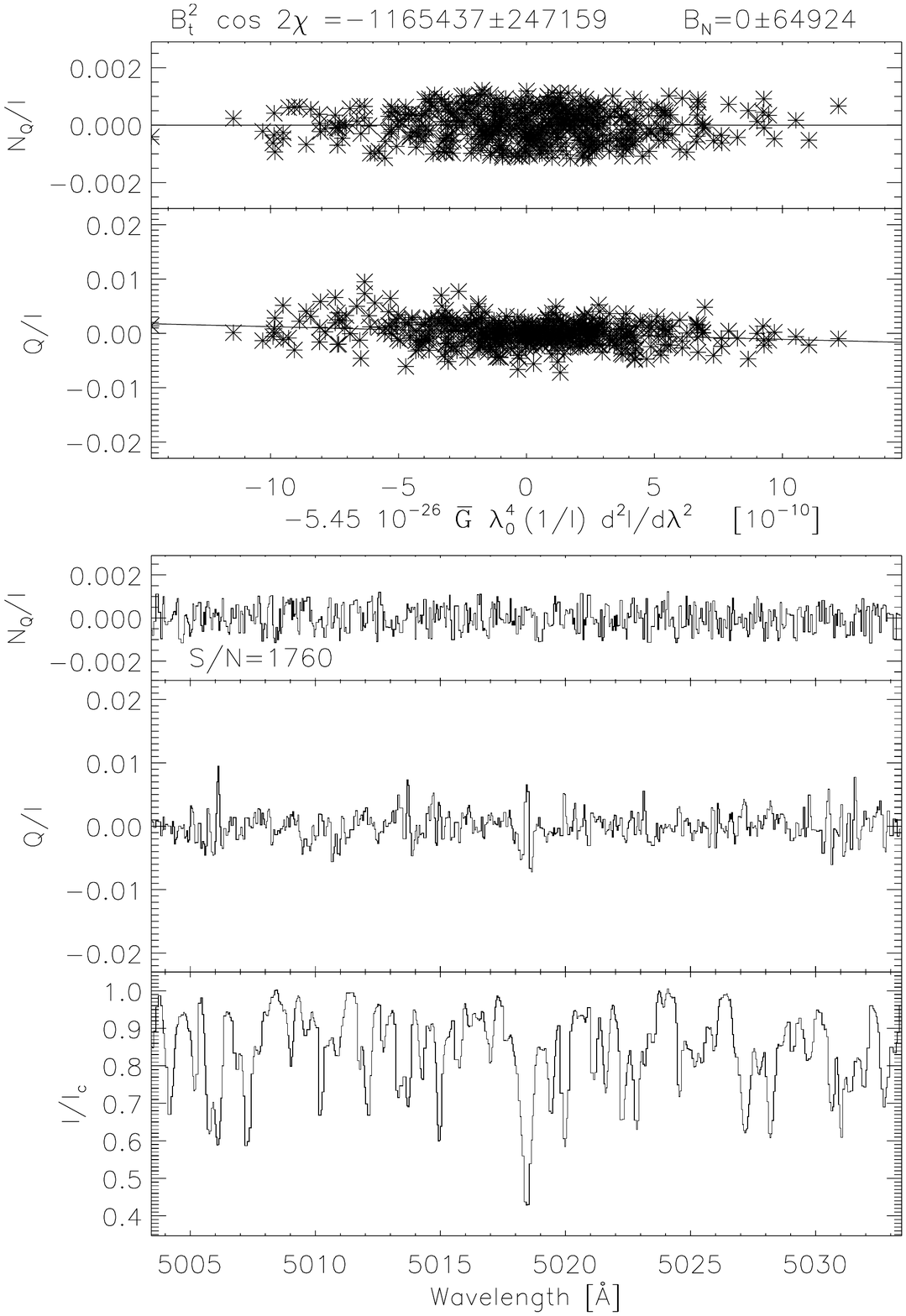}
\includegraphics[trim = 0.5cm 1.0cm 0.0cm 0.0cm, clip=true,width=5.9cm]{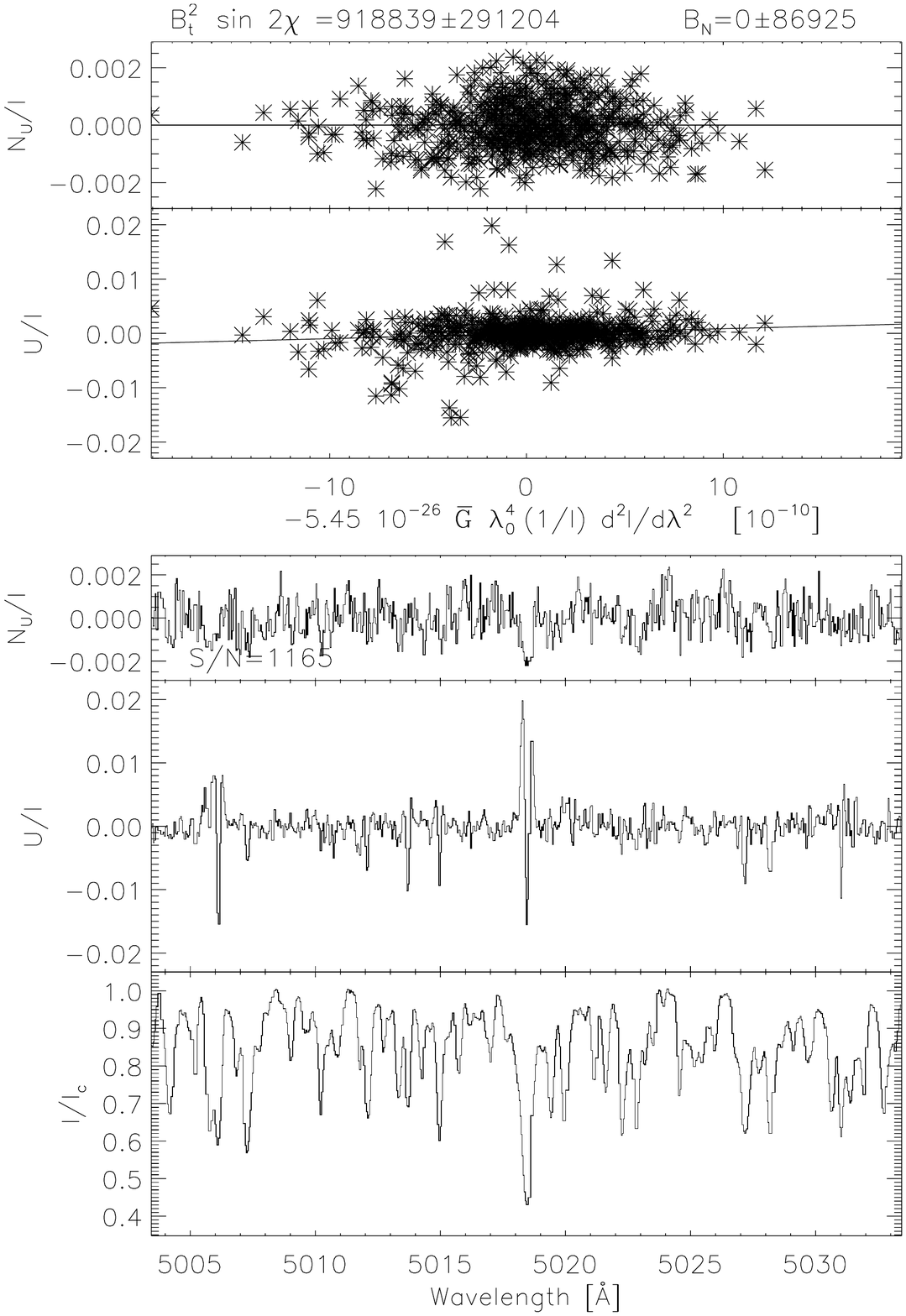}
\includegraphics[trim = 0.5cm 1.0cm 0.0cm 0.0cm, clip=true,width=5.9cm]{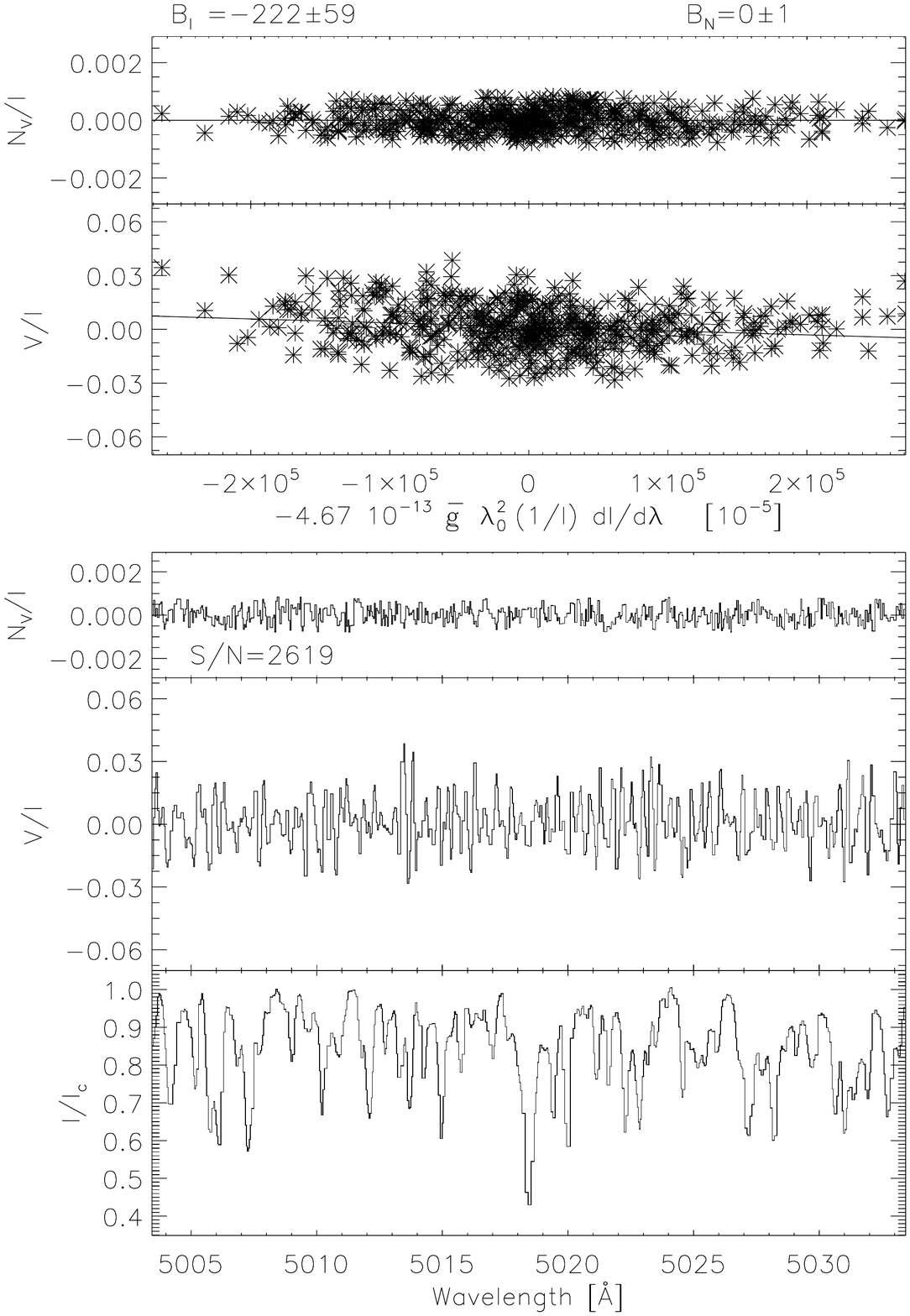}
\includegraphics[trim = 0.5cm 1.0cm 0.0cm 0.0cm, clip=true,width=5.9cm]{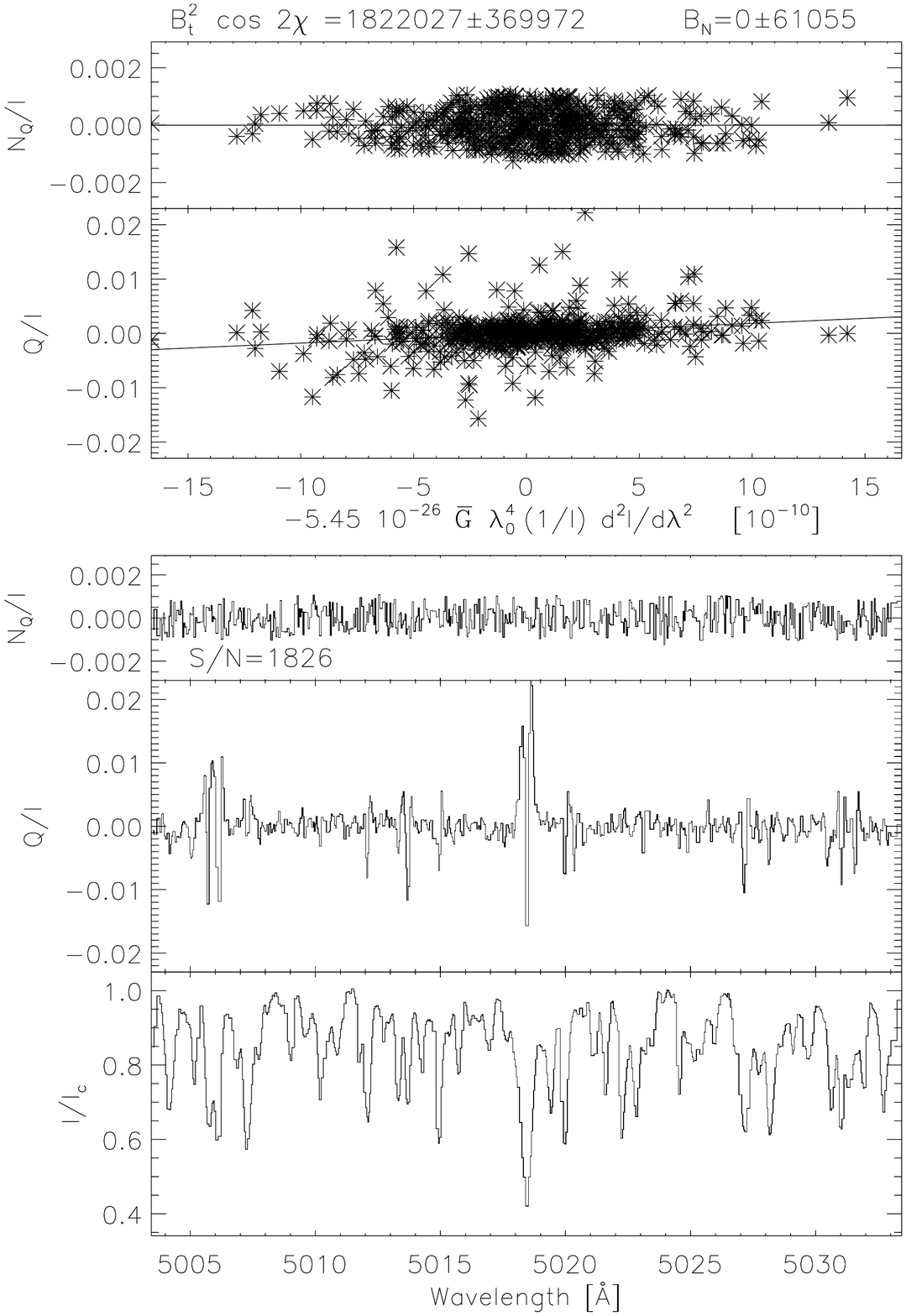}
\includegraphics[trim = 0.5cm 1.0cm 0.0cm 0.0cm, clip=true,width=5.9cm]{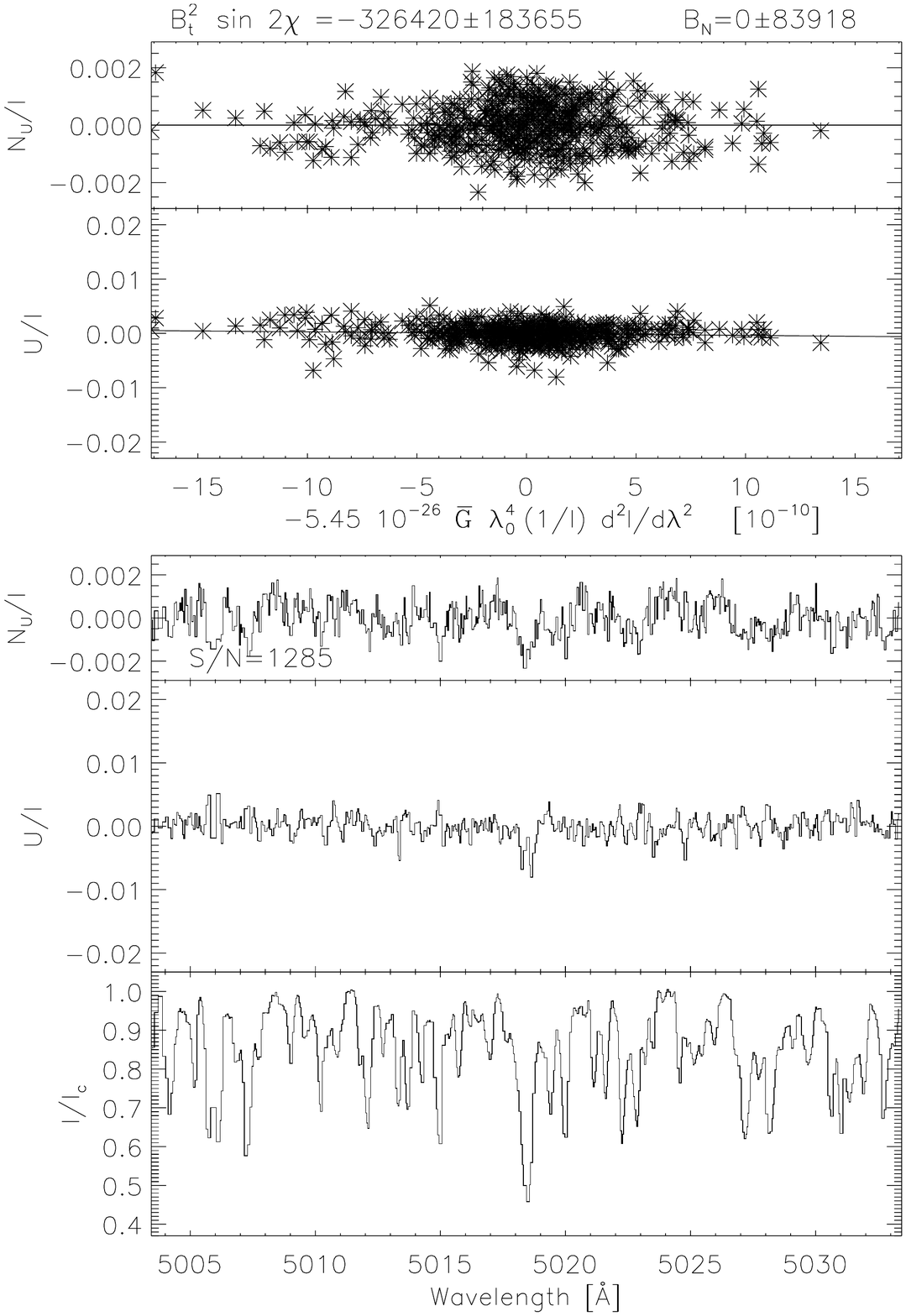}
\includegraphics[trim = 0.5cm 1.0cm 0.0cm 0.0cm, clip=true,width=5.9cm]{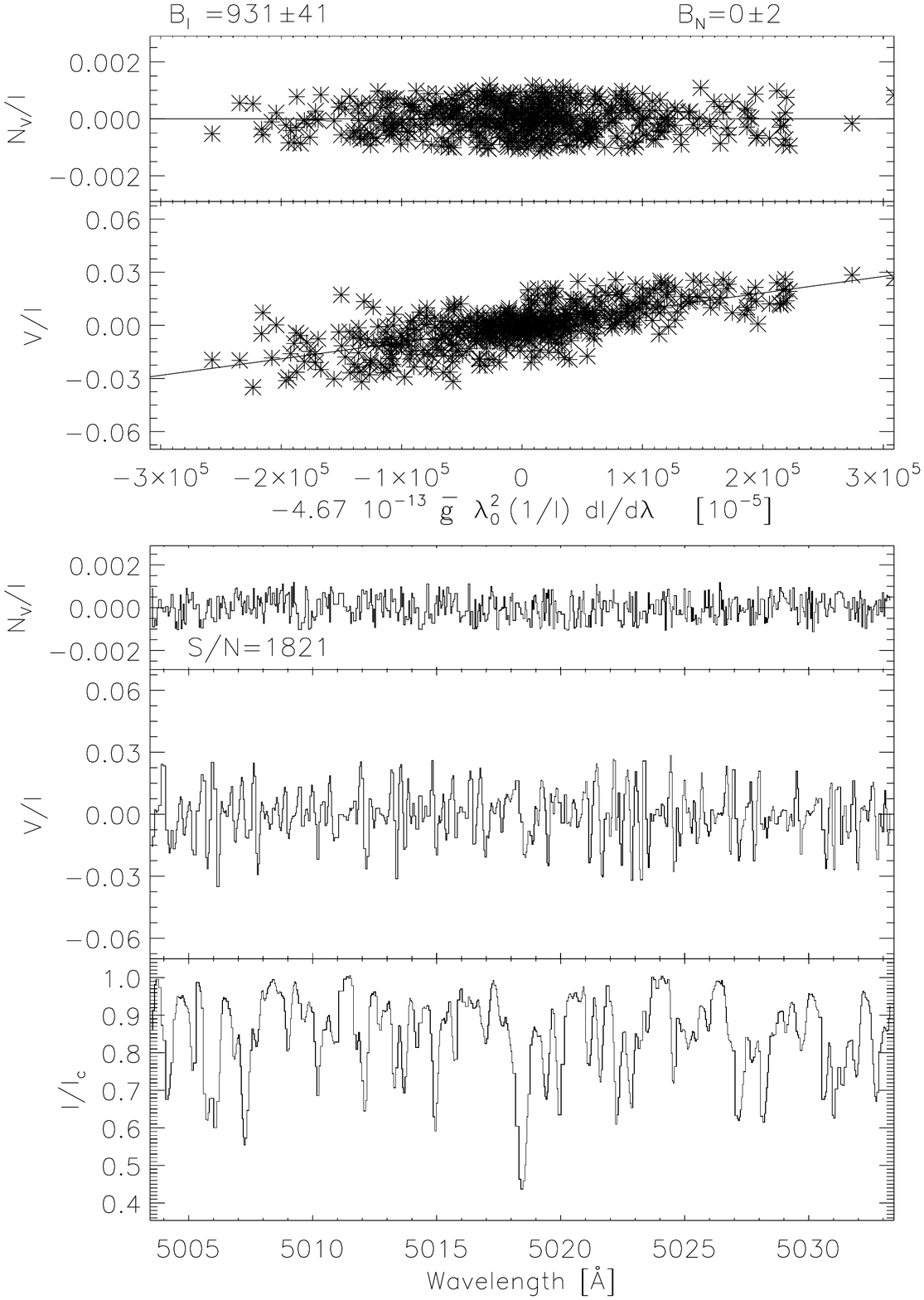}
\caption{Observed Stokes $I, Q, U$ and $V$ spectra of the magnetic star $\beta$\,CrB at
rotational phase $\phi=0.66$ (top block of six panels) and $\phi=0.85$ (bottom block of
six panels). Phase values are computed according to the ephemeris (\ref{eph}). Also
plotted are Stokes\,$Q$ and $U$ as a function of the second derivative of Stokes\,$I$,
 and Stokes\,$V$ as a function of the first derivative of Stokes\,$I$. {\it Noise}
spectra (Eq.\,\ref{Noise}) are shown to quantify the photon and extraction errors.
{\it Noise} constant with wavelength validates the correctness of the data reduction.
Slope with errors are reported.}
\label{Fig_slope}
\end{figure*}

\section{$\beta$\,CrB observations and data reduction}\label{Sec_Obs}

Ever since \cite{Babcock1949a}, $\beta$\,CrB has been one of the most
studied magnetic chemically peculiar main sequence star. Distinctive characteristics
of this class of stars are {\it a}) a very strong magnetic field as inferred from
the integrated Zeeman effect. Typical fields are  1 - 10\,kG, the strongest
known reaching $35$\,kG; {\it b}) Variability of the magnetic field, spectral
lines and luminosity with the same period\footnote{Periods typically measure
$2-10$\,d, however much shorter and longer periods have been found, see
\cite{Catalano1993} and references therein.}; the longitudinal magnetic field
often reverses its sign. So far, the oblique rotator is the only model that
provides an acceptable interpretation of the above-mentioned phenomena
\citep{Babcock1949b, Stibbs1950}. It is essentially based on two hypotheses:
{\it 1)} The magnetic field is largely dipolar with the dipole axis inclined
with respect to the the rotational axis, and {\it 2}) Over- and under-abundances
of chemical elements are distributed non-homogeneously over the stellar surface.
All observed variations are a direct consequence of stellar rotation.

For comparison with results on $\beta$\,CrB found in the literature we adopted
the measurements of the longitudinal field by \cite{Mathys1994}, the measurements
of the ``surface'' field (the integrated field modulus)
\begin{equation}B_s = \frac{1}{W \cal F_{\it I_{\rm c}}} \int_{0}^{2\pi}d\phi
\int_{0}^{\pi/2}|B|\cos\theta\,\sin\theta\,d\theta \times \int [I_{\rm c} - I_\lambda]d\lambda\label{Bs}
\end{equation}
by \cite{Mathys1997} and the ephemeris by \cite{Bagnulo01}:
\begin{equation}
JD(B_{||}^{\rm max}) = 2\,443\,310.221 + 18.4868\hspace{1cm} \rm days\label{eph}
\end{equation}

The linear polarization of $\beta$\,CrB has been measured 32 times over its
rotational period. These data have been obtained  with the {\it Catania Astrophysical
Observatory Spectropolarimeter} (CAOS) from June to July 2014 in the 370-860\,nm range with
resolution R = 55\,000 \citep{Leone16}, the minimum signal-to-noise ratio was S/N = 400.
With respect to the acceptance axis of the polarizer, we obtained Stokes\,$V$ by setting 
the fast axis of the quarter wave-plate retarder to $\alpha = +45\degr$ and
$-45\degr$ respectively. The fast axis of the half wave-plate retarder has been
rotated by $\alpha = 0\degr$ and $45\degr$ to measure Stokes\,$Q$, and by
$\alpha = 22.5\degr$ and $67.5\degr$ to measure Stokes\,$U$.

There are several methods to measure the degree of polarization from {\it o}-rdinary
and {\it e}-xtraordinary  beams  from the polarizer. As to the dual beam
spectropolarimetry, the ratio method was introduced by \cite{Tinbergen92}. It is
assumed that there is a time independent (instrumental) sensitivity $G$, for example
due to pixel-by-pixel efficiency variations -- together with a time dependent sensitivity $F$
of spectra -- for example due to variations in the transparency of the sky.
So a photon noise dominated Stokes parameter (generically $P  = V, Q$ or $U$)
can be obtained from the recorded {\it o}-rdinary and {\it e}-xtraordinary spectra,
  $S_{\alpha,o}$ and $S_{\alpha,e}$ respectively, at rotations $\alpha_1$ and $\alpha_2$ by:
\begin{equation}
\begin{aligned}
S_{\alpha_1,o} & = 0.5\,(I + P)\,G_{o} F_{\alpha_1} & S_{\alpha_1,e} & = 0.5\,(I - P)\,G_{e} F_{\alpha_1}  \\
S_{\alpha_2,o} & = 0.5\,(I - P)\,G_{o} F_{\alpha_2} & S_{\alpha_2,e} & = 0.5\,(I + P)\,G_{e} F_{\alpha_2} \nonumber
\end{aligned}
\end{equation}
Hence:
\begin{equation}
\frac{P}{I} = \frac{R_P - 1}{R_P + 1}\hspace{1.5cm}{\rm
with}\hspace{0.38cm}R_P^2 = \frac{S_{\alpha_1,o}/S_{\alpha_1,e}}{S_{\alpha_2,o}/S_{\alpha_2,e}} \nonumber
\end{equation}
In addition we compute the {\it noise} polarization spectrum:
\begin{equation}
\frac{N}{I} = \frac{R_N - 1}{R_N + 1}\hspace{1.5cm}{\rm with}\hspace{0.38cm}R_N^2
            = \frac{S_{\alpha_1,o}/S_{\alpha_2,e}}{S_{\alpha_2,o}/S_{\alpha_1,e}}\label{Noise}
\end{equation}
to check any possible error in Stokes $P/I$. Without errors, the {\it noise}
polarization spectrum is expected to present no dependence on the Stokes\,$I$
derivatives \citep{Leone07, Leone11}.

The preferred use of Eqs.\,\ref{gV}, \ref{gQ} and \ref{gU} over the original
relations given by \cite{Landi04} is due to the higher accuracy that can be
achieved in measuring $Q/I, U/I$ and $V/I$ as compared to $Q, U$ and $V$.

\section{Measuring magnetic field components}
\label{Sec_Reg}

As stated in the introduction, the linear fitting of Stokes\,$V$ versus the first
derivative of Stokes\,$I$ of Balmer line profiles has opened a new way to measure
$B_{||}$ of stars on the basis of low resolution spectra. Introducing this
method, \cite{Bagnulo02} quoted a series of papers based on photopolarimetry of
Balmer line wings to justify the validity of Eq.\,\ref{gV}
also for the whole visible disk of a star with a complex magnetic
field and despite the limb darkening \citep{Mathys2000}.

\begin{figure}
\centering
\includegraphics[trim = 1.2cm 1.2cm 2.0cm 1.0cm, clip=true, width=4.2cm]{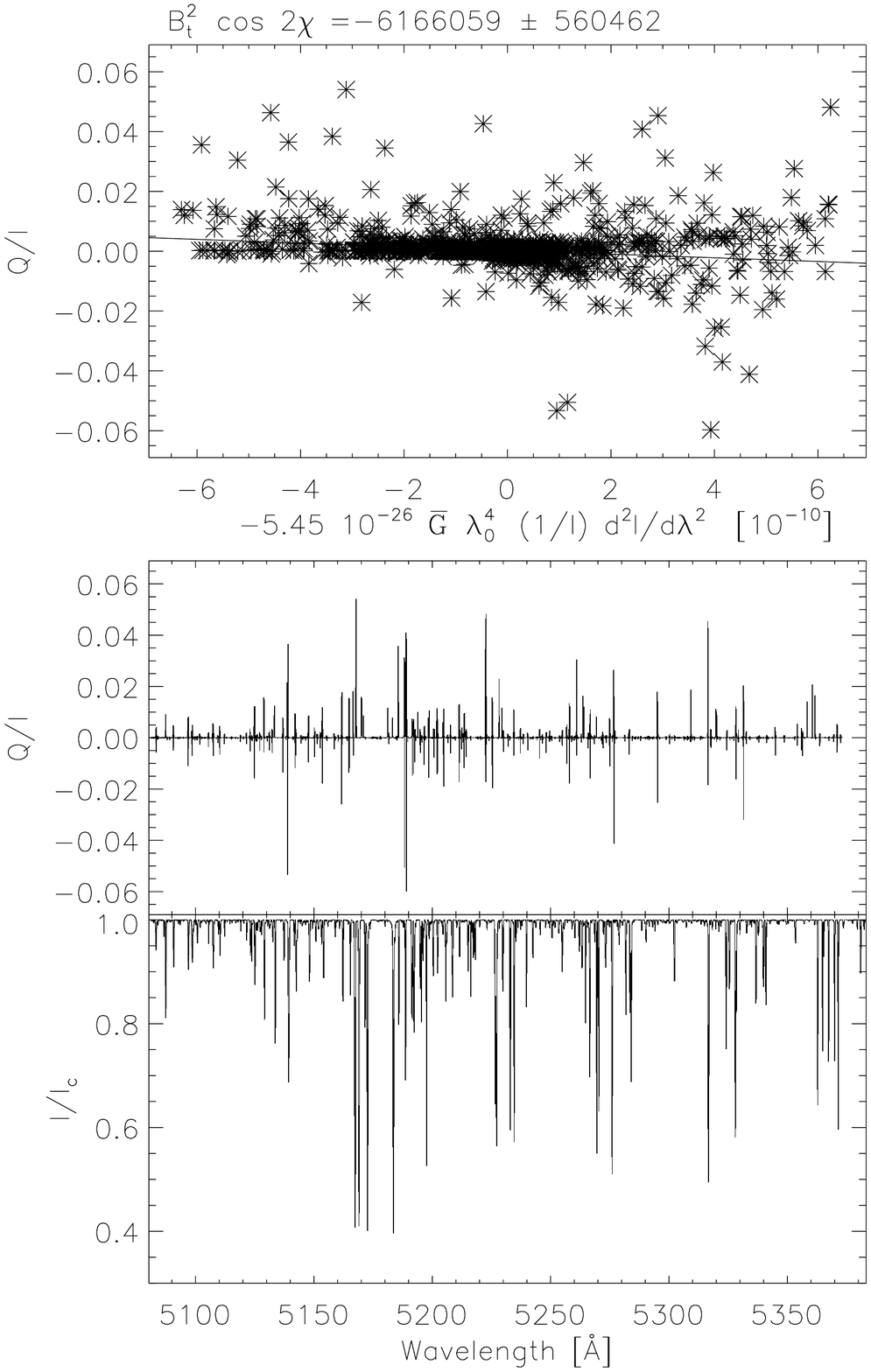}
\includegraphics[trim = 1.2cm 1.2cm 2.0cm 1.0cm, clip=true, width=4.2cm]{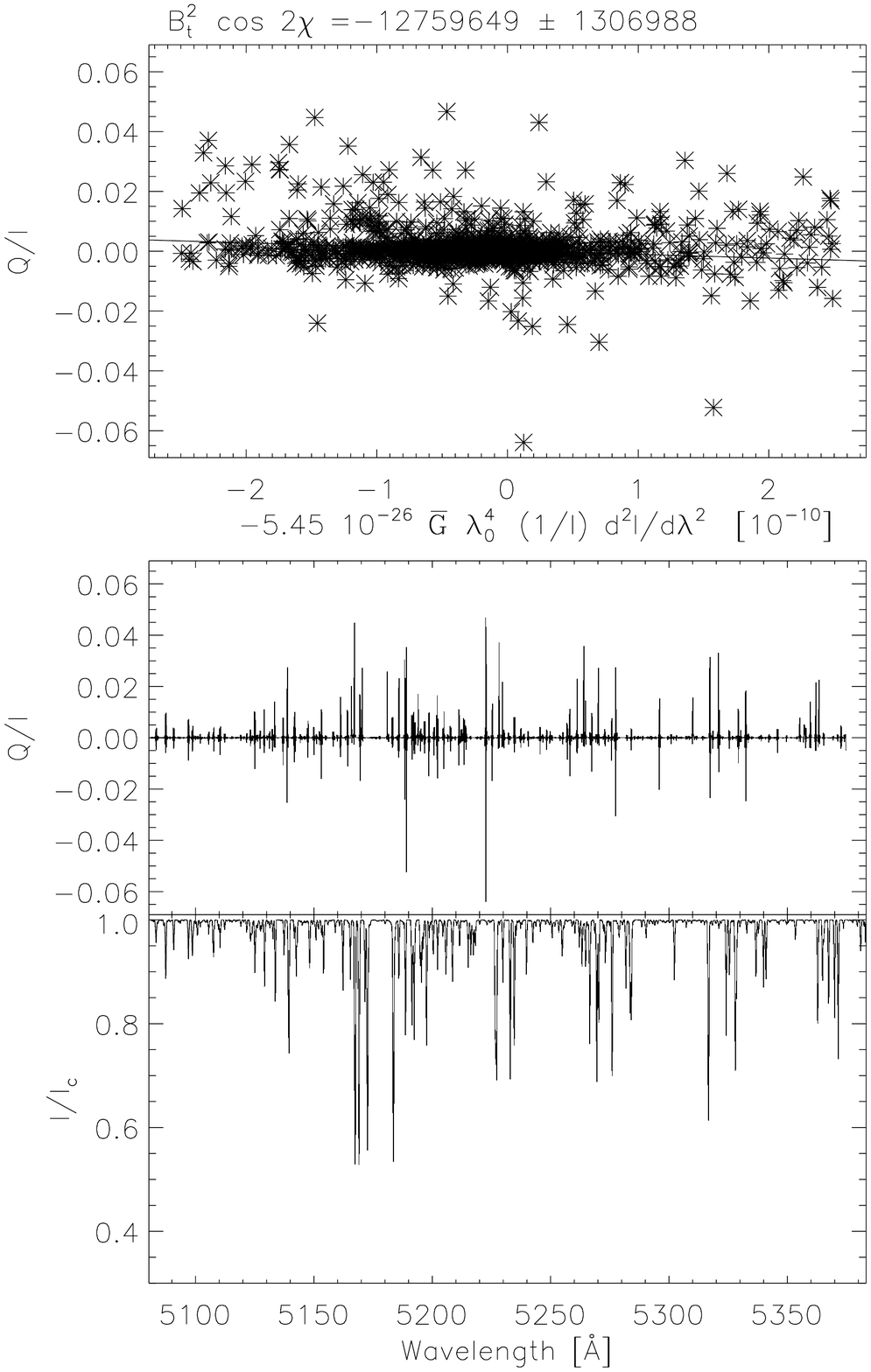}
\caption{Examples of {\sc Cossam} simulations for a magnetic dipole, B$_{\rm p}$ =  10\,kG,
orthogonal to the rotational axis and along the E-W direction. Left panel: star rotating
at 3\,km\,s$^{-1}$, right panel: 18\,km\,s$^{-1}$.}
\label{cossam:slope}
\end{figure}

\cite{Marian2012} have shown that Eqs.\,\ref{gV}, \ref{gQ} and \ref{gU} are valid
for disk-integrated line profiles of rotating stars with a magnetic dipolar field,
provided the rotational velocity is not larger than eight times the Doppler width of
the local absorption profiles. We have performed numerical tests with {\sc Cossam}
\citep{Stift2012} to find out how far the
derivative of the Stokes\,$I$ profile reflects Zeeman broadening before being dominated
by the rotational broadening. As a limiting case, we have assumed the dipole axis
orthogonal to the rotation axis, both being tangent to the celestial sphere. Two cases
are shown in Figure\,\ref{cossam:slope} and results are summarized in Table\,\ref{Tab:Simul}
for the spectral resolution of CAOS.
\begin{table}
\caption{\label{Tab:Simul} Ratio between the measured transverse field and the expected
value, derived by applying the derivative method. In this particular
case the equatorial velocity is equal to the projected velocity. }
\center
\begin{tabular}{cr|ccccc}
\hline
           & &\multicolumn{5}{c}{v$_{\rm eq}$ [km s$^{-1}$]} \\
                 &    & 0 & 3  & 6 & 12  &18  \\
\hline
\multirow{ 4}{*}{B$_p$[G]} &    10 & 1.93 & 1.18  & 1.17  & 0.79  & 3.51 \\
                           &    100 & 0.96 & 1.13  & 1.13  & 1.13  & 3.37 \\
                           &   1000 & 0.93 & 1.06  & 1.08  & 1.15  & 3.32 \\
                           &  10000 & 0.81 & 0.84  & 1.02  & 1.07  & 1.21 \\
\hline
\end{tabular}
\end{table}
These numerical simulations show that by applying the slope method, the transverse field
of a star observed with CAOS is estimated correctly to within 20\% for rotational
velocities up to 12\,km\,s$^{-1}$. We ascribe the anomalous value for a non-rotating star
with a weak (10\,G polar) field  to the fact that the spectral line profiles are dominated
by the 5.5 km\,s$^{-1}$ instrumental smearing.

We have also addressed the capability to measure the transverse component of fields
that are not purely dipolar. As a benchmark, we have extended the previous numerical tests with
{\sc Cossam} for a star rotating at 3\,km\,s$^{-1}$ and B$_{\rm p}$ = 10\,kG. The dipole, whose axis is still going
through the center of the star, has been displaced in the direction of the 
positive pole. As a function of the decentering in units of the stellar radius $a$,
the ratio between the mesured transverse field and the expected value
is  $r(a=0.0) = 0.84$, $r(a=0.1) = 1.06$, $r(a=0.2) = 1.16$, $r(a=0.3) = 1.15$,
and $r(a=0.4) = 4.57$

It appears therefore legitimate to apply the method to our spectra of $\beta$\,CrB,
which displays a rotational velocity of 3\,km\,s$^{-1}$ \citep{Ryabchikova2004}.

\subsection{The longitudinal field component of $\beta$\,CrB}
We have applied the method to our high resolution spectra and found a very high
precision of the measurements. Figure\,\ref{Fig_slope} shows Stokes\,$I$
and $V$ of $\beta$\,CrB at rotational phases $\phi = 0.66$ and 0.85 in a 30\,\AA\,
interval centered on the Fe\,{\sc ii}\,5018.44\,{\AA} line. Figure\,\ref{Fig_slope}
also shows Stokes\,$V$ as a function of the first derivative of Stokes\,$I$
and its linear fit.  If $\bar{g}$ = 1, the slope gives an error in the measured
$B_{||}$ of about 40\,G. It is worthwhile noting that the same procedure, as
applied to the {\it noise} spectra, gives a much smaller error of less than 4\,G.
We ascribe the 40\,G error to the line-by-line differences in the $\bar{g}$ Land\'e
factors, resulting in the superposition of straight lines with different slopes.
The observed Stokes $I$ and $V$ profiles of a generic spectral line $k$, with
effective Land\'e factor $g^k_{\rm eff}$, define a straight line in the 
$-4.67 \times 10^{-13}\lambda^{2}\frac{1}{I_\lambda}\frac{\partial I_\lambda}{\partial \lambda}$ vs $\frac{V_\lambda}{I_\lambda}$ plane
whose slope is $c_k = g^k_{\rm eff} B_{||}$. Using a set of $N$ spectral lines we
measure an average value for the longitudinal field $<B_{||}> = \frac{\sum c_k}{N} = <g_{\rm eff}> B_{||}$.
The relative error in the longitudinal field measure is given by the dispersion of the effective Land\'e factors.

Even though the precision is very high, the accuracy of the longitudinal field
measurements depends on the adopted $\bar{g}$ value; usually this is assumed equal
to unity. In \cite{Leone07}, we have numerically shown that the average value of
the Land{\'e} factors of the spectral lines of the magnetic star $\gamma$Equ,
observed in the 3780-4480\,{\AA} interval and weighted by their intensity, is about
1.1. As to $\beta$\,CrB, adopting the effective temperature, gravity and abundances
given by \cite{Ryabchikova2004}, we have extracted from VALD the list of expected
spectral lines and found an average value of $\bar{g} = 1.2 \pm 0.4$. We conclude
that the linear regression method measures the longitudinal field of a star with a
precision equal to the standard distribution of the effective Land{\'e} factors of
the spectral lines involved. 

\subsection{The transverse field component}
As an extension to the method described above to measure the longitudinal field,
we have plotted the Stokes\,$Q$ and $U$ signals as a function of the second
derivative of Stokes\,$I$ (Eqs.\,\ref{gQ} and \ref{gU}). Figure\,\ref{Fig_slope}
shows the expected linear dependencies for $\beta$\,CrB at two different rotational phases.

The conversion of the slopes to transverse field measures is less straightforward
than in the longitudinal case. Line-by-line differences in the second order Land{\'e}
factors are larger than differences in the effective Land{\'e} factors (Eq.\,\ref{foG}).
The second order Land{\'e} factors can become negative (Eq.\,\ref{soG}),
effective Land{\'e} factors only very exceptionally. In a list of solar Fe\,{\sc i} lines
given by \cite{Landi04} some 8\% of $\bar{G}$ values are negative.

Table \ref{Tab_Meas} reports the transverse field of $\beta$\,CrB by applying
Eqs.\,\ref{gQ} and \ref{gU} to 50\,{\AA} blocks of CAOS spectra in the 5000 to
6000\,{\AA} interval. As for $\bar{g}$, the adopted $\bar{G}$ of a block 
represents the average of the $G$ value of the predicted spectral lines. In order 
to check the reliability of our quantitative measurements of the transverse
field, we have applied the method also to the Fe\,{\sc ii}\,5018.44\,\AA\, line
which presents well defined Stokes profiles and is among the lines selected for
solar studies in the {\it T{\'e}lescope H{\'e}liographique pour l'Etude du
Magn{\'e}tisme et des Instabilit{\'e}s Solaires} (THEMIS). 

\begin{figure*}[h]
\begin{center}
\includegraphics[trim = 3.5cm 0.5cm 2.5cm 0.0cm, clip=true,width=8.cm]{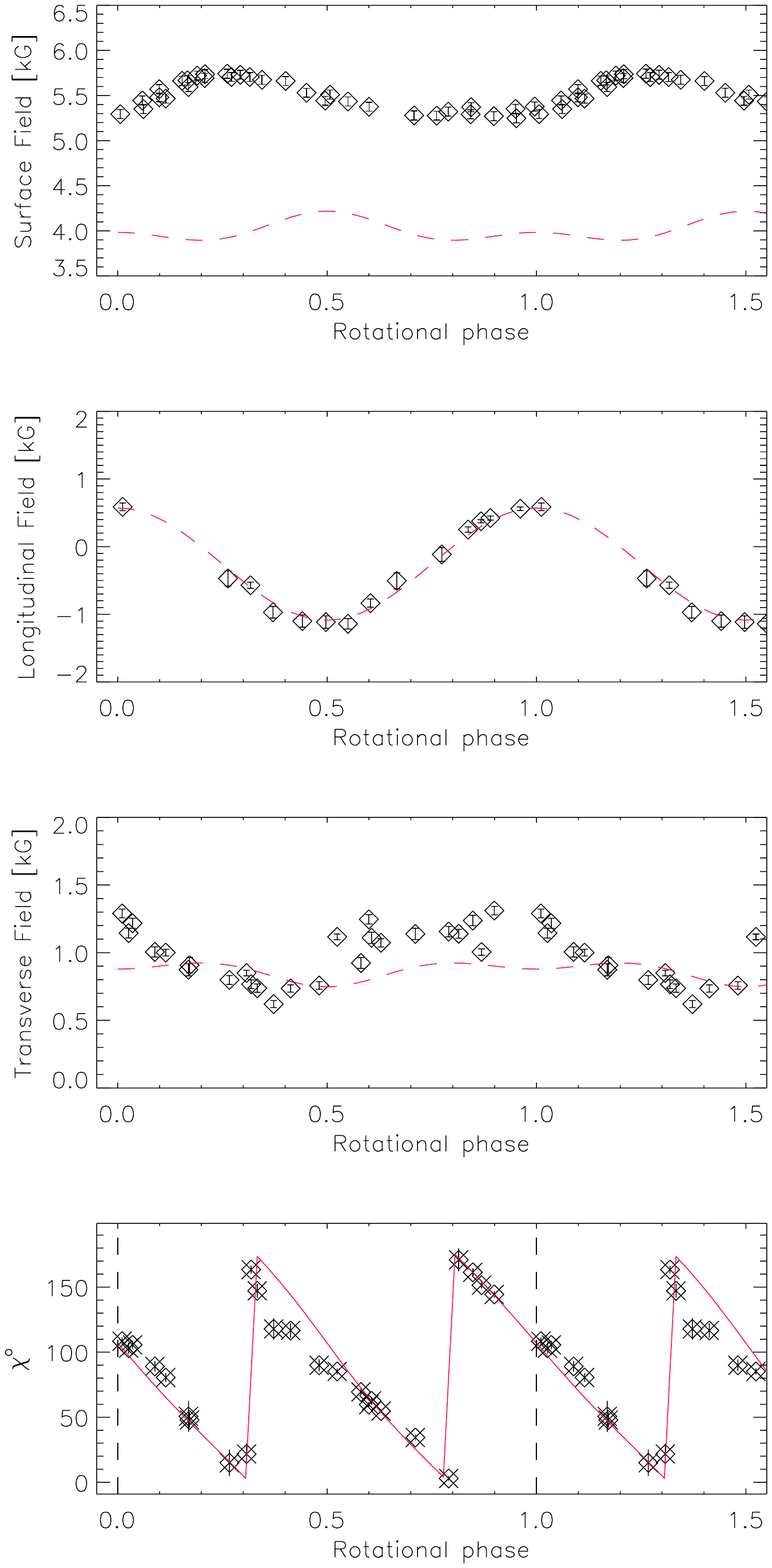}
\includegraphics[trim = 3.5cm 0.5cm 2.5cm 0.0cm, clip=true,width=8.cm]{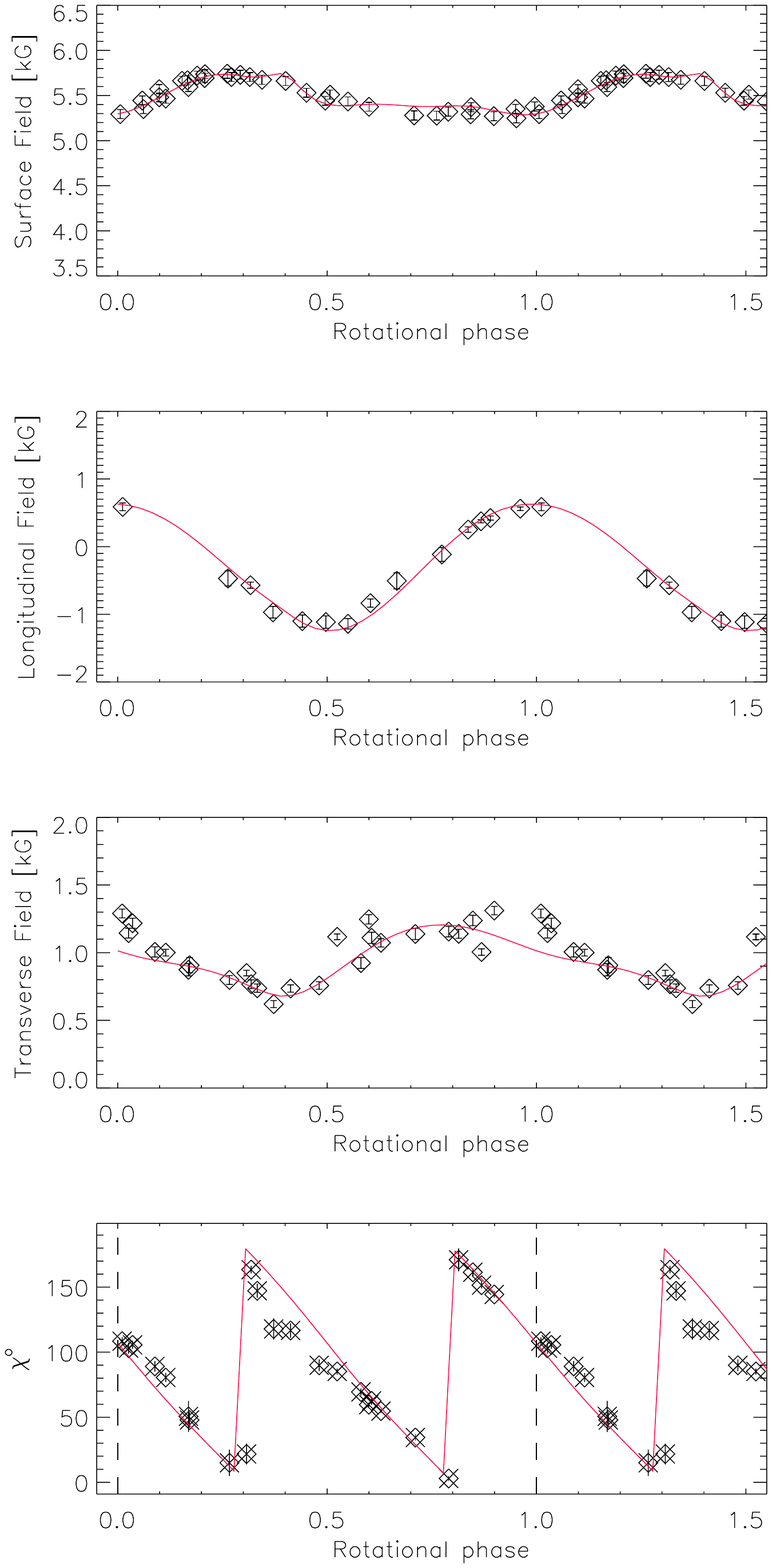}
\end{center}
\caption{Longitudinal and transverse magnetic field of $\beta$\,CrB as well as
the angle $\chi$ of the transverse component with respect to the North-South
meridian (measured counterclockwise) are plotted as a function of the rotational
phase. The left panels show the expected variations for a dipole, whereas the
right panels pertain to a field resulting from the superposition of a dipole,
a quadrupole and an octupole. The vertical line marks the positive extremum of
the longitudinal field, i.e. the rotational phase when the line of sight, the
rotation axis and the dipole axis all lie in the same plane. At this phase, the
transverse field is also aligned with the rotation axis; the angle $\chi$ gives
the orientation of the rotation axis with respect to the North-South direction
in the sky.}
\label{fig:HD137909}\end{figure*}

As applied to our collected spectra and on the basis of the ephemeris given in
Eq.\,\ref{eph}, $\beta$\,CrB presents a transverse field that varies with the
rotation period (Figure\,\ref{fig:HD137909}). The average value is about
1\,kG and the amplitude as large as 0.25\,kG. A comparison (Table \ref{Tab_Meas})
with results from the Fe\,{\sc ii}\,5018.44\,\AA\, line reveals general agreement;
however the associated errors are larger. We suppose that the error in measuring the
transverse field -- i.e. the slope error -- is dominated by the scatter in the second
order Land\'e factors, similar to what we found for the longitudinal field.  

The angle $\chi$ is variable too with the rotation period, see Figure\,\ref{fig:HD137909}.
Since by definition, $\chi$ is limited to the range $0 - 180\degr$, it exhibits a saw-tooth
behavior.

\begin{table}
\center
\caption{\label{tab:Bt} Measured transverse magnetic field of $\beta$\,CrB.
Eqs. \ref{gQ} and \ref{gU} have been applied to CAOS spectra in the range 5000 to 6000 \AA\, and to a well known single iron line.}
\begin{tabular}{c|rr|rr}
\hline
    &   \multicolumn{2}{c}{5000 - 6000 \AA} & \multicolumn{2}{|c}{Fe{\sc ii}\,5018.44\,\AA}   \\\hline
HJD  &   \multicolumn{1}{c}{ $B_{\bot}\pm\sigma$ }    & \multicolumn{1}{c|}{  $\chi\pm\sigma$ }  &\multicolumn{1}{c}{$B_{\bot}\pm\sigma$}    & \multicolumn{1}{c}{$\chi\pm\sigma$ } \\
2450000  &   \multicolumn{1}{c}{kG} & \multicolumn{1}{c|}{\degr}  &  \multicolumn{1}{c}{kG} & \multicolumn{1}{c}{\degr}  \\
\hline 
  6787.511 &  1.117$\pm$0.084 &  85$\pm$ 5 &  0.930$\pm$0.012   &    88$\pm$ 1 \\
  6788.560 &  0.923$\pm$0.160 &  69$\pm$ 5 &  0.877$\pm$0.015   &    70$\pm$20\\
  6799.481 &  0.907$\pm$0.223 &  48$\pm$12 &  0.841$\pm$0.015   &    45$\pm$43\\
  6802.471 &  0.737$\pm$0.123 & 147$\pm$11 &  0.564$\pm$0.025   &   148$\pm$31\\
  6807.515 &  1.109$\pm$0.162 &  63$\pm$ 6 &  0.897$\pm$0.019   &    63$\pm$27\\
  6809.448 &  1.138$\pm$0.167 &  34$\pm$ 5 &  0.982$\pm$0.016   &    37$\pm$35\\
  6815.452 &  1.217$\pm$0.147 & 106$\pm$ 6 &  11207$\pm$0.019   &   101$\pm$13\\
  6816.436 &  1.005$\pm$0.149 &  89$\pm$ 8 &  0.995$\pm$0.015   &    84$\pm$ 5 \\
  6820.484 &  0.849$\pm$0.082 &  22$\pm$16 &  0.794$\pm$0.021   &    12$\pm$ 9 \\
  6822.429 &  0.735$\pm$0.108 & 117$\pm$11 &  0.608$\pm$0.022   &   127$\pm$40\\
  6826.417 &  1.073$\pm$0.122 &  55$\pm$ 3 &  0.960$\pm$0.014   &    55$\pm$36\\
  6829.409 &  1.155$\pm$0.153 &   3$\pm$ 4 &  1.010$\pm$0.013   &     1$\pm$ 1 \\
  6830.474 &  1.237$\pm$0.155 & 161$\pm$ 6 &  1.094$\pm$0.015   &   160$\pm$20\\
  6831.422 &  1.311$\pm$0.124 & 144$\pm$ 5 &  1.152$\pm$0.013   &   143$\pm$37\\
  6833.478 &  1.289$\pm$0.129 & 108$\pm$ 6 &  1.134$\pm$0.020   &   108$\pm$20\\
  6835.405 &  1.000$\pm$0.098 &  81$\pm$ 7 &  0.949$\pm$0.019   &    75$\pm$14\\
  6836.408 &  0.873$\pm$0.180 &  51$\pm$24 &  0.703$\pm$0.019   &    44$\pm$41\\
  6844.369 &  1.246$\pm$0.136 &  60$\pm$ 3 &  1.030$\pm$0.014   &    60$\pm$30\\
  6848.338 &  1.139$\pm$0.134 & 171$\pm$17 &  1.000$\pm$0.014   &   173$\pm$ 7 \\
  6849.343 &  1.004$\pm$0.094 & 152$\pm$ 5 &  0.976$\pm$0.021   &   150$\pm$29\\
  7129.556 &  1.144$\pm$0.142 & 103$\pm$ 9 &  1.129$\pm$0.014   &    96$\pm$ 7 \\
  7189.467 &  0.798$\pm$0.130 &  15$\pm$20 &  0.691$\pm$0.017   &     5$\pm$ 3 \\
  7190.426 &  0.766$\pm$0.128 & 163$\pm$ 6 &  0.733$\pm$0.020   &   169$\pm$12\\
  7191.416 &  0.619$\pm$0.102 & 118$\pm$17 &  0.483$\pm$0.024   &   135$\pm$41\\
  7193.425 &  0.757$\pm$0.112 &  90$\pm$ 9 &  0.657$\pm$0.018   &    93$\pm$ 7 \\
\hline
\end{tabular}\label{Tab_Meas}
\end{table}

\section{The added-value of the transverse field}
Large efforts have gone into the study of stellar magnetic fields \citep{Mestel1999}
but it is still not possible to predict the magnetic field geometry of an ApBp star.
As mentioned in the introduction, the magnetic variability of early-type upper main
sequence stars is thought to be due to a mainly dipolar field, with the dipole axis
inclined with respect to the rotational axis. Once the mean field modulus could be
determined in addition to the longitudinal field it became clear that the magnetic
configurations went beyond simple dipoles \citep{Preston1967}. \cite{Deutsch70} was the
first to model the field with a series of spherical harmonics, \cite{Landstreet1970} 
introduced the decentred dipole, \cite{Landstreet2000} adopted a field characterized by
a co-linear dipole, quadrupole and octupole geometry and \cite{Bagnulo2002b}
modeled the field by a superposition of a dipole and a quadrupole field, arbitrarily
oriented.

It has been known for quite some time that the surface field of $\beta$\,CrB cannot
be represented by a simple dipole \citep{Wolff70}. Let us however, for the present purpose, look at the variability of the longitudinal 
field within the framework of a pure dipole \citep{Stibbs1950}
\begin{equation}
B^d_{||}(t) =\frac{15 + u}{15 -5 u}  \frac{B_d}{4}\left(\cos i\cos\beta_d +\sin i\sin\beta_d\cos\frac{2\pi}{P}\right) \label{Stibbs}
\end{equation}
-- where $u$ is the limb coefficient, $i$ the angle between the line of sight and
the rotation axis, $\beta_d$ the angle between dipole and rotation axes, $B_d$
the magnetic field strength at the poles and $P$ the rotation period.
Hence, the \cite{Schwarzschild1950} relation 
\begin{equation}
B^d_{||}(min, max) = \frac{15 + u}{15 -5 u} \frac{B_d}{4} \cos(\beta_d \pm i) \label{Schw}
\end{equation}
and \cite{Preston1971} relation
\begin{equation}
\tan i\,\tan\beta_d = \frac{1 - r_{||}}{1 + r_{||}} \label{Prest}
\end{equation}

-- where $r_{||}$ is the ratio between minimum and maximum longitudinal field values --
one can establish combinations of $i$, $\beta_d$ and $B_d$ which match an observed
sinusoidal $B_{||}$ variability. We note that the combination $i = 153\degr$,
$\beta_d = 81\degr$ and $B_d=6.2$\,kG yields the observed average value of the
transverse field, however underestimating the surface field (left panel of
Figure\,\ref{fig:HD137909}). On the other hand, adopting $i = 161\degr$,
$\beta_d = 84\degr$ and $B_d = 8.6$\,kG, we obtain a match for the average field
modulus, but now the transverse field is overestimated.

In order to correctly predict the observed variability of longitudinal, transverse 
and surface field of $\beta$\,CrB, it is obviously necessary to assume a magnetic field
geometry without cylindrical symmetry \citep{Mathys1993}. We have thus decided
to model the magnetic variability by taking a dipole, a quadrupole and an octupole
with symmetry axes pointing in different directions with respect to the rotation
axis and with respect to each other. As the reference plane we adopt the plane
defined by the rotation axis and the line of sight; the rotation phase $\phi$ is
zero when the dipole axis lies in this plane. The right panel of
Figure\,\ref{fig:HD137909} shows the result of our best fit with
$i = 153\degr$,\\
$B_d = \,\,+6.1$\,kG,\, $\beta_d = 86\degr$,\\
$B_q = \,\,+3.9$\,kG,\, $\beta_q = 88\degr$,\, $\phi_q = 64\degr$,\\
$B_o = -10.7$\,kG,\,  $\beta_o = 20\degr$,\, $\phi_o = 40\degr$.\\
$\phi_q$ and $\phi_o$ represent the azimuth of quadrupole and octupole
respectively.

The problem of the uniqueness of this particular magnetic configuration is outside the
scope of this paper. At present we focus exclusively on the added value of knowing
the transverse component in relation to the orientation of the rotational axis, the
radius and the equatorial velocity of magnetic stars.

\begin{figure}
\begin{center}
\includegraphics[trim = 3.5cm 0.5cm 2.5cm 0.0cm, clip=true,width=4.2cm]{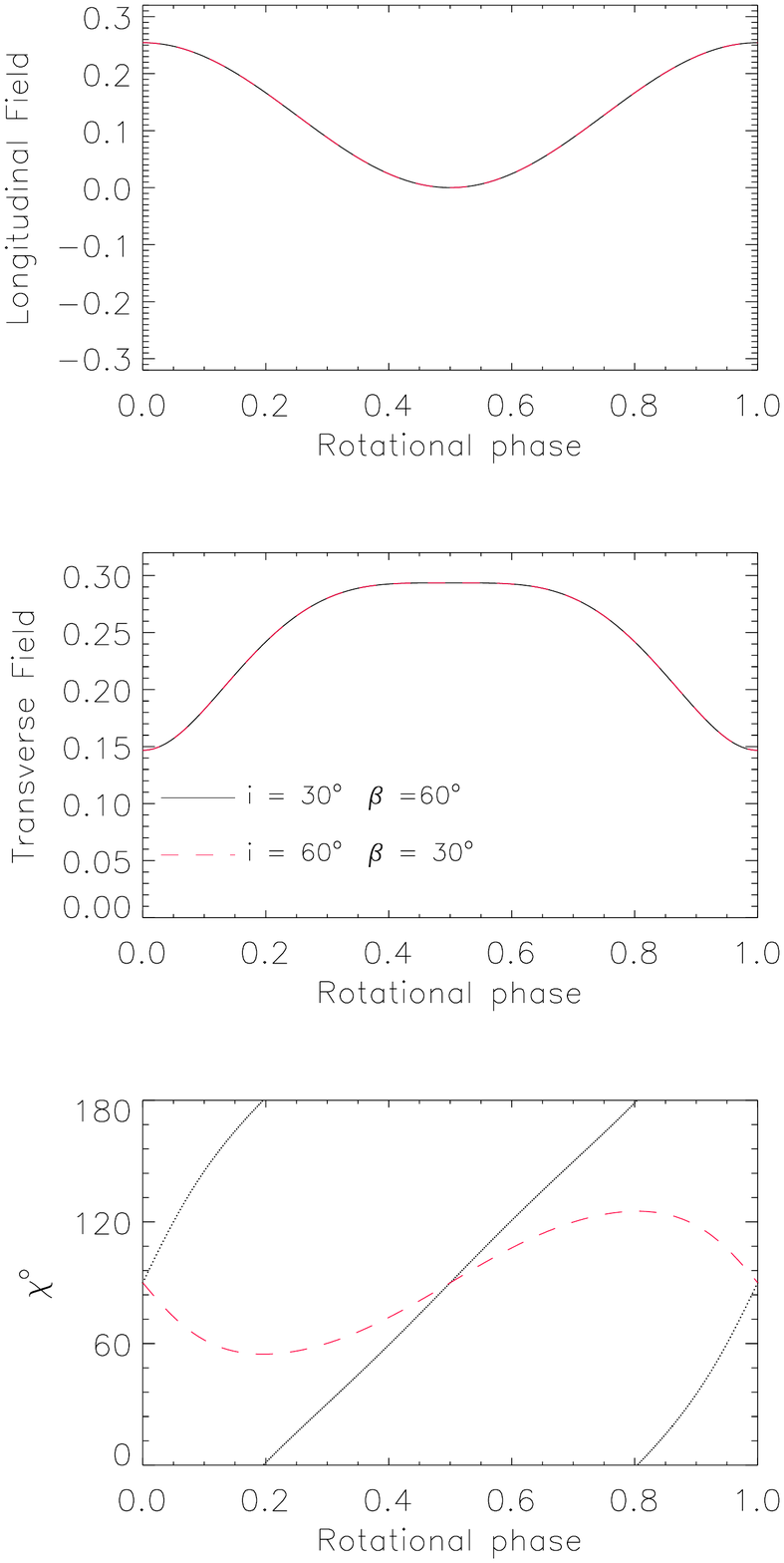}
\includegraphics[trim = 3.5cm 0.5cm 2.5cm 0.0cm, clip=true,width=4.2cm]{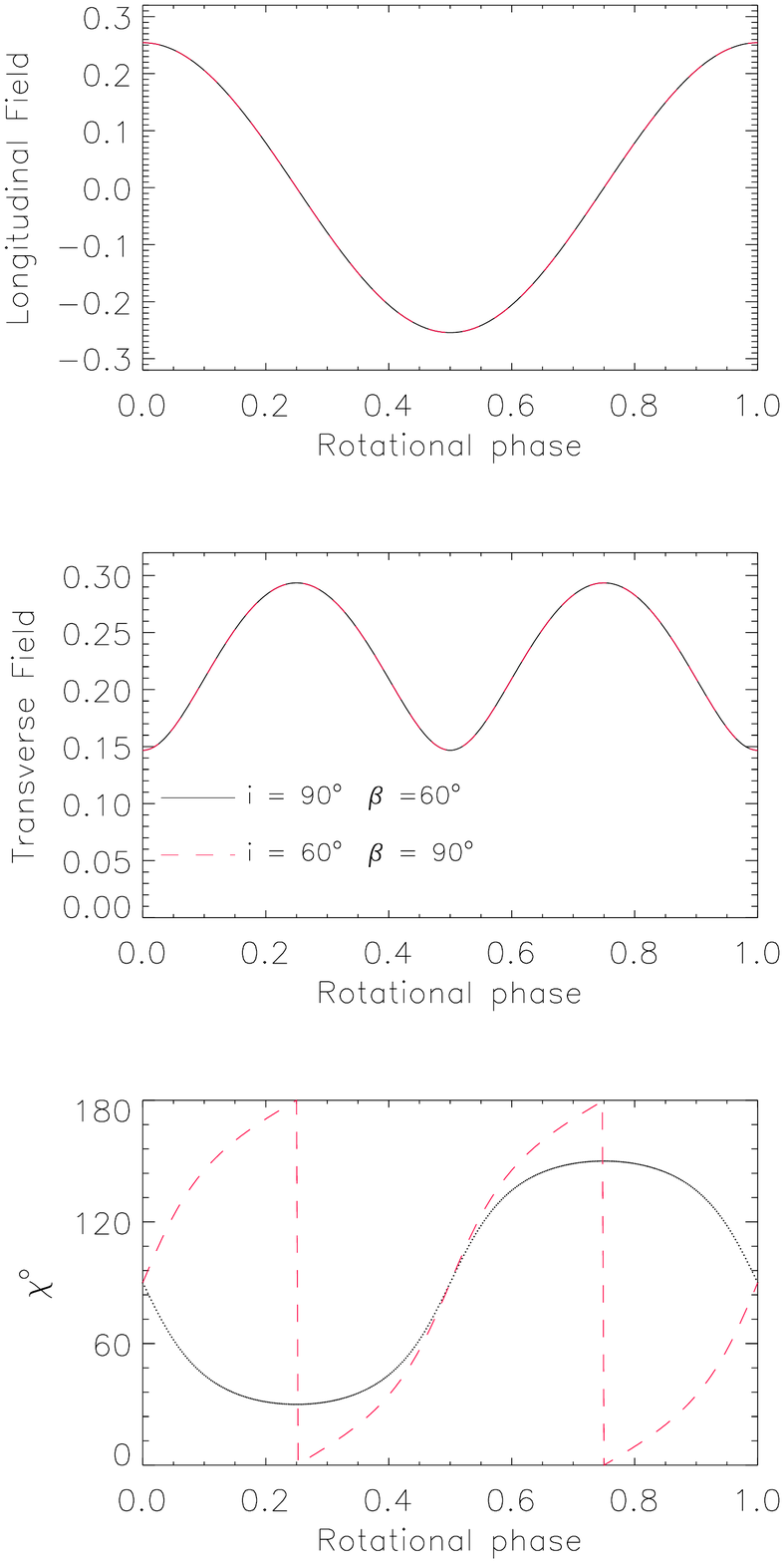}
\end{center}
\caption{Two examples of magnetic dipoles indistinguishable from the respective
longitudinal field variations that present very different $\chi$ variations.
Field values are given in units of the polar strength.}
\label{fig:disentangle}\end{figure}

\subsection{Degeneracy between $i$ and $\beta_d$}
Fig.\,\ref{fig:HD137909} shows that the angle $\chi$ is dominated by the dipolar
component with only a negligible dependence on the higher order components of the
magnetic field. This doesn't really come as a surprise: \cite{Schwarzschild1950}
has shown that the maximum value of the longitudinal field is equal to $\sim\,$30\% 
of the polar value for a dipole and equal to $\sim$\,5\% for a quadrupole.
Numerical integration over the visible stellar disk shows that the
same holds true for the transverse field:
$B_{||}^{max} = B_{\perp}^{max}  \,\sim 0.30\,B_d \sim 0.05\,B_q \sim 0.02\,B_o$,
considering also the octupole.
This is an intuitive result since the longitudinal field for an observer
simply is the transverse field of another observer located at 90\degr. For example,
the longitudinal component as measured by an observer located above the north pole of a dipole is
the transverse component for an observer lying in the magnetic equator. The latter
can see half of the southern hemisphere that presents exactly the magnetic field
configuration of the invisible half of the north hemisphere.

It is straight to show that the previous relations \ref{Stibbs}, \ref{Schw} and \ref{Prest} together with an equal set of relations  where $i$ is replaced by
$i + 90\degr$, that are valid for the transverse field, break the degeneracy between $i$ and $\beta_d$. We conclude that
the knowledge of the transverse field component removes the indeterminacy in the Schwarzschild relation (Eq.\,\ref{Schw}) between the
angles formed by the rotation axis with the  line-of-sight ($i$) and the magnetic axis ($\beta$).

We note that it is not necessary to solve these equations to solve the degeneracy between $i$ and $\beta_d$ when the $\chi$ variation 
with the stellar rotation is available. It happens that, if $i$ is larger than $\beta_d$ the $\chi$  variation is not a sawtooth (Figure\,\ref{fig:disentangle}).

\subsection{Orientation of the stellar rotational axis}\label{Sec_Orient}
The longitudinal and transverse components of a dipolar field are projected
along the dipole axis. This, within the framework of the oblique rotator model,
describes a cone around the rotation pole. It happens that when we observe the
extrema of the longitudinal field, the transverse field is projected onto the
rotation axis. This means that, when we observe the extrema of the longitudinal
field, the measured angle $\chi$ represents the angle between the rotation axis
and the North-South direction in the sky. This simple consideration gives us
the possibility to determine the absolute orientation of the rotation axis of
a star hosting a dominant dipolar magnetic field. From our data we conclude
that the rotation axis of $\beta$\,CrB is tilted by about 110$\degr$ with
respect to the N-S direction.

\subsection{Equatorial velocity and Stellar radius}

Once the degeneracy between $i$ and $\beta_d$ removed, the stellar radius can
be inferred from the relation valid for a rigid spherical rotator
\begin{equation}
\rm v_{\rm e}\,\sin i [km\,s^{-1}]\, P[days] = 50.6\,\, R[R_{\odot}] \sin i
\end{equation}
where $P$ is the rotational period. As to $\beta$\,CrB, \cite{Kurtz2007} report a
$\rm v_{\rm e}\,\sin i$ in the range $3.0 - 3.8$\,km\,s$^{-1}$. The indeterminacy
($i =  153\degr, \beta_d = 81\degr$) or ($i =  81\degr, \beta_d = 153\degr$)
from the Schwarzschild relation would thus result in the following values of
the stellar radius: $1.25\pm 0.15\, R_{\odot}$ or $2.75\pm 0.30\, R_{\odot}$. Our
determination of the angle $i = 153\degr$ (implying $2.75\, R_{\odot}$) agrees
with the interferometric value of $2.6\, R_{\odot}$ for the radius of $\beta$\,CrB
obtained by \cite{Bruntt2010}. The equatorial velocity lies between 6.6 and
8.4 km\,s$^{-1}$.

\section{Conclusions}

The linear regression between Stokes\,$V$ and the first derivative
of Stokes\,$I$ in low resolution spectroscopy was introduced by
\cite{Bagnulo02} as a method for estimating the longitudinal magnetic
fields of faint stars.

We have carried out phase-resolved and high-resolution full Stokes
spectropolarimetry of the magnetic chemically peculiar star $\beta$\,CrB with
the {\it Catania Astrophysical Observatory Spectropolarimeter} \citep{Leone16}.
On the basis of these data, we have shown that it is possible to extend the
previous method to the high resolution spectropolarimetry with the more general
aim of recovering the Stokes profiles hidden in the photon noise. A condition
of faint stars as observed at low resolution but also of very weak stellar
magnetic fields. The precision appears to be limited by our knowledge
of Land\'e factors and by the non homogeneous distribution of
chemical elements on the visible disk. \cite{Leone04} found that measuring the
longitudinal field, element by element, different values are obtained monitoring
the equivalent width variations with the rotation period of HD\,24712.

We have also shown that a regression of Stokes\,$Q$ and $U$ with respect to
the second derivative of Stokes\,$I$ provides a direct measure of the transverse
component of a stellar magnetic field and its orientation in the sky. If the
magnetic field is not symmetric with respect to the rotation axis, the transverse
field vector rotates in the sky. Having found that the dipolar component of the
field is mainly responsible for the transverse component, we conclude that it is
possible to determine the orientation of the rotation axis with respect to the
sky: the value of the angle between the rotation axis and the North-South direction
corresponds to the value of $\chi$ at the rotational phase where the longitudinal
field reaches an extremum, viz. $\Theta = \chi(B_{||}^{extrem.})$

To our knowledge, the transverse component has never before been measured directly.
The interpretation of broadband linear photopolarimetry by \cite{Landi1981}, based
on the linear polarization properties of spectral lines formed in the presence of a
magnetic field and its application to phase-resolved data by \cite{Bagnulo1995} to
constrain the magnetic field geometries of chemically peculiar stars represent an
approach somewhat similar to ours. It is worthwhile noting that $\beta$\,CrB has been
modeled from phase-resolved broadband linear photopolarimetry by \cite{Leroy1995}
and by \cite{Bagnulo2000} who found $\Theta = 135\degr$ and $\Theta = 124\degr$
respectively. These values have to be compared with our result of $\Theta = 110\degr$.

In view of the improving capability to obtain high resolution spatial observations
via optical and radio interferometry, it becomes increasingly important to know
the orientation of the rotation axis in the sky. The determination of the transverse
field is thus fundamental in multi-parametric problems such as the 3D mapping of
the magnetospheres of early-type radio stars \citep{Trigilio2004, Leone2010, Trigilio2011}.

\bibliographystyle{aasjournal}
\bibliography{ms.bib}

\end{document}